\documentclass{emulateapj}
\journalinfo{To Appear in the Astrophysical Journal}

\newcommand{\kt}{\ensuremath{k_{\rm{B}}T}}
\newcommand{\fx}{\ensuremath{F_{\rm{X}}}}
\newcommand{\lx}{\ensuremath{L_{\rm{X}}}}
\newcommand{\nh}{\ensuremath{N_{\rm{H}}}}
\newcommand{\ve}{\ensuremath{V_{\rm{EM}}}}

\newcommand{\zoxy}{\ensuremath{Z_{\rm{O}}}}
\newcommand{\zneo}{\ensuremath{Z_{\rm{Ne}}}}

\slugcomment{}
\shorttitle{X-ray Fading and Expansion of the GK Persei Remnant}
\shortauthors{D. Takei et al.}

\begin{document}
%%%%%%%%%%%%%%%%%%%%%%%%%%%%%%%%%%%%%%%%%%%%%%%%%%%%%%%%%%%%
\title{X-ray Fading and Expansion in the ``Miniature Supernova Remnant'' of GK Persei}
\author{
D.~Takei\altaffilmark{1},
J.~J.~Drake\altaffilmark{2},
H.~Yamaguchi\altaffilmark{3,4},
P.~Slane\altaffilmark{2},
Y.~Uchiyama\altaffilmark{5}, \&
S.~Katsuda\altaffilmark{6}
}
\email{takei@spring8.or.jp}
\altaffiltext{1}{Institute of Physical and Chemical Research (RIKEN), RIKEN SPring-8 Center,
                 1-1-1 Kouto, Sayo, Hyogo 679-5148, Japan}
\altaffiltext{2}{Smithsonian Astrophysical Observatory, 60 Garden Street, Cambridge, MA 02138, USA}
\altaffiltext{3}{NASA Goddard Space Flight Center, Code 662, Greenbelt, MD 20771, USA}
\altaffiltext{4}{Department of Astronomy, University of Maryland, College Park, MD 20742, USA}
\altaffiltext{5}{Department of Physics, Rikkyo University,
                 3-34-1 Nishi-Ikebukuro, Toshima, Tokyo 171-8501, Japan}
\altaffiltext{6}{Japan Aerospace Exploration Agency, Institute of Space and Astronautical Science,
                 3-1-1 Yoshino-dai, Chuo-ku, Sagamihara, Kanagawa 252-5210, Japan}
%%%%%%%%%%%%%%%%%%%%%%%%%%%%%%%%%%%%%%%%%%%%%%%%%%%%%%%%%%%%

\begin{abstract}
 We report on a second epoch of \textit{Chandra} X-ray imaging spectroscopy of the
 spatially-resolved old nova remnant GK Persei. An ACIS-S3 observation of 97.4~ks
 was conducted in November 2013 after a lapse of 13.8 years from the last visit in
 2000. The X-ray emitting nebula appeared more faint and patchy compared with the
 first epoch. The flux decline was particularly evident in fainter regions and the
 mean decline was 30--40\% in the 0.5--1.2~keV energy band. A typical expansion of
 the brightest part of the remnant was 1$\farcs$9, which corresponds to an expansion
 rate of 0$\farcs$14~yr$^{-1}$. The soft X-ray spectra extracted from both the 2000
 and 2013 data can be explained by a non-equilibrium ionization collisional plasma
 model convolved with interstellar absorption, though do not allow us to constrain
 the origin of the flux evolution. The plasma temperature has not significantly
 evolved since the 2000 epoch and we conclude that the fading of the X-ray emission
 is due largely to expansion. This implies that recent expansion has been into a
 lower density medium, a scenario that is qualitatively consistent with the structure
 of the circumstellar environment photographed soon after the initial explosion more
 than a century ago. Fainter areas are fading more quickly than brighter areas,
 indicating that they are fainter because of a lower ambient medium density and
 consequently more rapid expansion.
\end{abstract}


%%%%%%%%%%%%%%%%%%%%%%%%%%%%%%%%%%%%%%%%%%%%%%%%%%%%%%%%%%%%
\section{Introduction}\label{sc:intro}
%%%%%%%%%%%%%%%%%%%%%%%%%%%%%%%%%%%%%%%%%%%%%%%%%%%%%%%%%%%%
Structures resulting from stellar explosions represent space laboratories that elucidate
the physics of enormous blasts under extreme conditions, and are of fundamental
importance to astrophysics. The representative of this kind of structure is a supernova
remnant (SNR).  X-ray observations of SNRs have provided wide-ranging insights into
their plasma physics, shock phenomena, and into Cosmic-ray origins (e.g.,
\citealt{slane1999,uchiyama2007,yamaguchi2009d,katsuda2013x}). However, the time history
of an individual SNR is generally difficult to track because of the long timescales ---
comparable to the entirety of human written history --- over which typical remnants
evolve. With the notable exception of spectacular nearby remnants, such as Cas~A
\citep[e.g.,][]{Fesen.etal:06,Patnaude.Fesen:09}, our understanding of remnant expansion
and evolution has therefore largely depended on theory, numerical simulations, and
systematic snapshot observational assessments of objects sampling different evolutionary
phases.

This paper presents a different approach to studying the X-ray evolution of an explosion
remnant. We study not a supernova but a ``classical'' nova resulting from a cataclysmic
explosion in an accreting binary system comprising a white dwarf and a red dwarf or
giant companion. The released energy and material propagate through the circumstellar
environment similar to, but on much smaller scales than, those in supernovae. Classical
novae are about 10$^{5}$ times smaller in energy, and typically factors of
$10^{5}$--$10^{7}$ times smaller in ejecta mass. The analogy with supernovae renders a
nova remnant a miniature SNR with a short life: its full dynamical evolution occurs on a
timescale comparable with a human lifetime and can potentially be followed in X-rays by
multi-epoch imaging spectroscopy.

Our target is Nova Persei 1901 (hereafter GK Per), which hosts the largest known X-ray
emitting nebula centered on a white dwarf binary \citep{seaquist1989}.  In the
following, we start by comparing \textit{Chandra} Cycle-1 data of GK Per obtained in
2000 with our recent observation in Cycle-15 obtained in 2013. The earlier data have
been analyzed in detail by \citet{balman2005} who characterized the shock conditions,
energetics of the remnant, and the shocked mass. Here, we use the second epoch data to
examine the remnant expansion, cooling and fading, and discuss the results in the
context of the physics of expanding explosion remnants.

%%%%%%%%%%%%%%%%%%%%%%%%%%%%%%%%%%%%%%%%%%%%%%%%%%%%%%%%%%%%
\section{Target (GK Persei)}\label{sc:target}
%%%%%%%%%%%%%%%%%%%%%%%%%%%%%%%%%%%%%%%%%%%%%%%%%%%%%%%%%%%%
GK~Per is a magnetic white dwarf binary at the distance of 470~pc \citep{mclaug1960s}
that underwent a classical nova outburst on 1901 February 21 \citep{williams1901o}.
Light echoes from the event photographed in 1901 and 1902 illuminated a complex of
shell-like structures around the central object \citep{Ritchey:02,Perrine:02} that
\citet{Bode.etal:87} and \citet{Bode.etal:04} interpreted as part of an old planetary
nebula. This nebula is also apparent as a vaguely bipolar-shaped structure, elongated
in the northwest-southeast direction, in the light of H$\alpha$, [\ion{O}{3}]~$\lambda$
5007, and the \textit{IRAS} 100~$\mu$m band.
\citet{Schaefer:88} noted that the lack of the sort of reflection features seen in
GK~Per around other novae indicates that the grain density around GK~Per is unusually
high --- in keeping with the planetary nebular origin suggested by \citet{Bode.etal:87}.

The extended nova remnant and its evolution has been imaged extensively in the optical
over several decades \citep[e.g.,][]{Seaquist.etal:89,Anupama.Prabhu:93,Slavin.etal:95,
Lawrence.etal:95,shara2012g,liimets2012t}. It has a pronounced asymmetric appearance,
with brighter emission to the southwest and a dearth of emission in the east. The limb
defined by the extent of optical knots exhibits a decidedly flattened appearance to the
northwest and northeast.
{\it Hubble Space Telescope} (\textit{HST}) observations in 1995 and 1997 revealed the
details of the remnants complex structure comprising a myriad of knots \citep{shara2012g}.
Its present-day angular size based on optical images from 2011 December 13 is about
1$\arcmin$ in radius, and the knots are observed to be expanding with circular symmetry
\citep{liimets2012t}. \citet{anupama2005} found a possible deceleration of the ejecta
after 1950 for the period 1901--2003, suggesting that the remnant is in a transition to
an adiabatic phase. A typical expansion rate of the outermost optical knots in
2004--2011 was 0$\farcs$3--0$\farcs$5~yr$^{-1}$, corresponding to a velocity of
600--1000~km~s$^{-1}$ \citep{liimets2012t}, which is similar to the 1995--1997 results
found by \cite{shara2012g}.

The various properties of the remnant were discussed in detail by \citet{seaquist1989}.
The long axis of the bipolar \ion{H}{1} and \textit{IRAS} 100~$\mu$m map lies in the
northwest-southeast direction, and essentially orthogonal to the northeast-southwest
axis (of symmetry of the X-ray remnant, see below). \citet{seaquist1989} concluded
that the nova ejecta are interacting with a cone of this pre-existing ambient medium
in the southwest where the remnant appears brightest. This region also corresponds
to a particularly bright structure lying at an angular separation of approximately
0$\farcs$5--1$\arcmin$ from the central star in the 1901 and 1902 images by
\citet{Ritchey:02} and \citet{Perrine:02}. \citet{seaquist1989} suggested this might
be a cone of material ejected during the planetary nebular phase suggested by
\citet{Bode.etal:87} and that the remnant is being decelerated by interaction with
this material. \citet{seaquist1989} also showed that the remnant expansion was
consistent with an adiabatic evolution in a medium with density declining with the
inverse square of the radial distance. Here the \citet{seaquist1989} analysis was
based on data obtained up until the late 1980's, and \citet{anupama2005} discussed
subsequent observations made until 2003, presenting evidence that the deceleration
of the blast expansion had decreased since 1987. They concluded that the density
into which the southwest quadrant was expanding in the later data was lower that in
1987 by a factor of about 7.

Regarding non-optical wavelengths, non-thermal radio emission indicating the presence
of accelerated electrons interacting with the remnant magnetic field was detected in
the southwest quadrant by \citet{Reynolds.Chevalier:84}, who estimated that a minimum
of 1\% of the 10$^{45}$~erg explosion energy has gone into the non-thermal electrons
and the magnetic field.
In the X-ray regime, GK~Per has the only nova remnant for which clumps were spatially
resolved. \textit{ROSAT} discovered the extended nebula for the first time in 1996
\citep{balman1999}, finding the bulk of the X-ray emission came from the southwestern
quadrant, then \textit{Chandra} revealed the presence of local X-ray structures in a
``lumpy and asymmetric nebula'' in 2000 \citep{balman2005}. The \textit{Chandra} ACIS
spectrum was dominated by a softer optically-thin thermal plasma component below
1.0--1.2~keV, but a harder non-thermal component was also found by \cite{balman2005},
who concluded that the remnant was still in the adiabatic phase in which the radiated
energy loss is still negligible in comparison with the initial explosion energy.

%%%%%%%%%%%%%%%%%%%%%%%%%%%%%%%%%%%%%%%%%%%%%%%%%%%%%%%%%%%%
\section{Observations and Reduction}\label{sc:obs}
%%%%%%%%%%%%%%%%%%%%%%%%%%%%%%%%%%%%%%%%%%%%%%%%%%%%%%%%%%%%
\textit{Chandra} has now visited GK~Per twice for imaging spectroscopy using the ACIS-S3
back-illuminated CCD. The first exposure was taken on 2000 February 10 (Obs.ID $=$ 650)
with a total observing time of 95.3~ks. Our additional observation was obtained on 2013
November 22 (Obs.ID $=$ 15741) with an exposure of 97.4~ks (see Table~\ref{tb:param} for
details). Both data sets were re-processed by the standard pipeline using the
\textit{Chandra} Interactive Analysis of Observations (CIAO) package \citep{frusci2006c}
with calibration database version 4.6.2.

\begin{table}[tb]
 \begin{center}
  \caption{Observation Logs and Spectral Properties\tablenotemark{a}}\label{tb:param}
  \begin{tabular}{llll}
   \tableline
                                            &                                      & \multicolumn{2}{c}{Obs.ID}                      \\
   Par.                                     & Unit                                 & 650                    & 15741                  \\
   \tableline
   $\textit{t}$\tablenotemark{b}            & (d)                                  & 36148.0                & 41182.2                \\
   $\textit{t}$\tablenotemark{b}            & (yr)                                 & 99.0                   & 112.8                  \\
   $\textit{t}\rm{_{str}}$\tablenotemark{c} & (UT)                                 & 2000-02-10             & 2013-11-22             \\
                                            &                                      & 01:15:21               & 20:51:14               \\
   $\textit{t}\rm{_{end}}$\tablenotemark{c} & (UT)                                 & 2000-02-11             & 2013-11-24             \\
                                            &                                      & 04:46:38               & 00:44:32               \\
   $\textit{t}\rm{_{exp}}$\tablenotemark{c} & (ks)                                 & 95.3                   & 97.4                   \\
   $\theta\rm{_{off}}$\tablenotemark{d}     & (arcmin)                             & 1.25                   & 0.18                   \\
   \tableline
   SC\tablenotemark{e,f}                    & (counts)                             & 3693                   & 793                    \\
   NR\tablenotemark{e,f}                    & (counts s$^{-1}$)                    & 3.06$\times$10$^{-2}$  & 5.68$\times$10$^{-3}$  \\
   ME\tablenotemark{e,f}                    & (keV)                                & 0.61                   & 0.66                   \\
   HR\tablenotemark{g}                      &                                      & $-$0.68$\pm$0.10       & $-$0.73$\pm$0.10       \\
   \tableline
   \nh                                      & (10$^{21}$ cm$^{-2}$)                & 2.10 (fixed)           & 2.10 (fixed)           \\
   \kt                                      & (keV)                                & 0.23$_{-0.04}^{+0.03}$ & 0.22$_{-0.05}^{+0.05}$ \\
   \zoxy                                    & (solar)                              & 0.31$_{-0.06}^{+0.09}$ & 0.16$_{-0.05}^{+0.09}$ \\
   \zneo                                    & (solar)                              & 2.84$_{-1.01}^{+1.55}$ & 1.71$_{-0.93}^{+1.32}$ \\
   $\tau$                                   & (10$^{9}$ s cm$^{-3}$)               & 4.82$_{-1.26}^{+5.57}$ & 3.77$_{-1.57}^{+6.89}$ \\
   \ve\tablenotemark{h}                     & (10$^{54}$ cm$^{-3}$)                & 1.18$_{-0.22}^{+0.48}$ & 1.63$_{-0.62}^{+1.50}$ \\
   \tableline
   \fx\tablenotemark{f}                     & (10$^{-14}$ ergs~s$^{-1}$~cm$^{-2}$) & 8.09$_{-0.96}^{+1.00}$ & 4.84$_{-1.28}^{+1.56}$ \\
   \lx\tablenotemark{f,h}                   & (10$^{30}$ ergs~s$^{-1}$)            & 9.70$_{-1.15}^{+1.20}$ & 5.76$_{-1.53}^{+1.86}$ \\
   \tableline
   \multicolumn{2}{l}{$\chi^{2}/\rm{d.o.f.}$}                                      & 38.4/28                & 10.7/14                \\
   \tableline
  \end{tabular}
  \begin{flushleft}
  $^{\rm{a}}${Statistical uncertainties indicate 1$\sigma$ confidence ranges.} \\
  $^{\rm{b}}${Elapsed days and years from the discovery of the nova (15436.6~d
  in modified Julian date) to the middle of each observation.} \\
  $^{\rm{c}}${Start dates ($\textit{t}\rm{_{str}}$), end dates
  ($\textit{t}\rm{_{end}}$), and exposure times ($\textit{t}\rm{_{exp}}$).} \\
  $^{\rm{d}}${Off-axis angles ($\theta\rm{_{off}}$) of the white dwarf binary,
  of which the nova explosion in 1901 caused the surrounding remnant.} \\
  $^{\rm{e}}${Source counts (SC), net rates (NR), and median energies (ME).} \\
  $^{\rm{f}}${Values are estimated in the 0.5--1.2~keV energy band.} \\
  $^{\rm{g}}${Hardness ratios (HR) of the remnant spectra. The values were
  defined as (H$-$S)/(H$+$S), where H and S are photon fluxes in the harder
  (0.8--1.2~keV) and softer (0.5--0.8~keV) bands, respectively.} \\
  $^{\rm{h}}${Values are for a distance of 470~pc \citep{mclaug1960s}.} \\
  \end{flushleft}
 \end{center}
\end{table}

\clearpage
%%%%%%%%%%%%%%%%%%%%%%%%%%%%%%%%%%%%%%%%%%%%%%%%%%%%%%%%%%%%
\section{Analysis}\label{sc:ana}
%%%%%%%%%%%%%%%%%%%%%%%%%%%%%%%%%%%%%%%%%%%%%%%%%%%%%%%%%%%%
\subsection{Diffuse Emission and Comparison}
%%%%%%%%%%%%%%%%%%%%%%%%%%%%%%%%%%%%%%%%%%%%%%%%%%%%%%%%%%%%
As found by \citet{balman2005} from the 2000 February data, the latest observation
reveals the brightest X-ray emission to lie in the southwest quadrant and to be
characterized by a prominent arc with extended ``wings'' of emission running from the
northwest to the southeast. The source is notably fainter than it appeared in 2000.

\begin{figure}[tb]
 \epsscale{1.00}
 \plotone{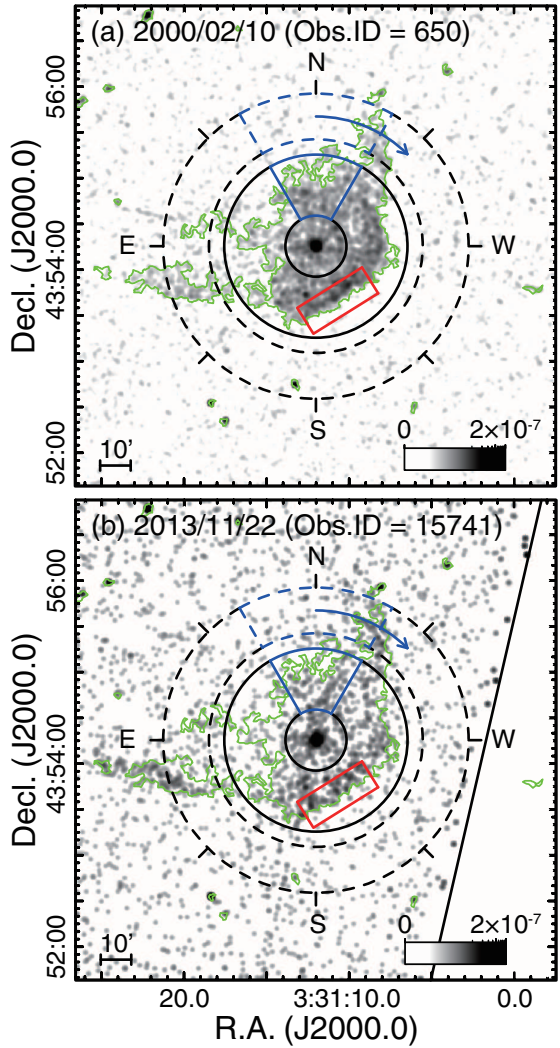}
 \caption{\textit{Chandra} ACIS-S3 0.5--1.2~keV photon flux images of GK Per in (a) 2000
 and (b) 2013. Data were smoothed by a Gaussian blur with a 2$\sigma$ radius of 5~pixels
 (2$\farcs$46), and were rendered with a logarithmic intensity greyscale covering up to
 2$\times$10$^{-7}$~photons~cm$^{-2}$~s$^{-1}$. The source and background extraction
 regions are indicated by the solid and dashed annuli, respectively. Areas of diffuse
 emission identified by the \texttt{vtpdetect} algorithm are shown as green
 polygons. The 45$^{\circ}$ blue sectors show the source and background extraction
 regions that were rotated in a clockwise direction to investigate the azimuthal
 dependence of the X-ray flux (see text and Figure~\ref{fg:az}).  The red rectangles of
 size 50$\arcsec\times$20$\arcsec$ parallel to a position angle of 122$\degr$ are the
 regions used for the projection profiles of Figure~\ref{fg:proj}. On the lower image,
 the edge of the CCD chip is also shown by the solid black line.
 }\label{fg:im}
\end{figure}

We limit detailed analysis to soft X-ray events in the 0.5--1.2~keV energy band.
At lower energies, the detector efficiency, especially around the carbon K-edge at
0.28~keV, is significantly attenuated by a layer of molecular contamination that
appears to be depositing on the instrument filter \citep{marshall2004c,chandra2013u}.
At higher energies, the source fluxes become comparable to, or lower than, the
background levels. The background-subtracted X-ray count rates were estimated to be
3.06$\pm$0.07$\times$10$^{-2}$ and 5.68$\pm$0.32$\times$10$^{-3}$~counts~s$^{-1}$
in 2000 and 2013, respectively, corresponding to a decline of 81$\pm$1\%. Based on
folding a model fit to the observed spectrum (see below) through the instrument
effective areas in the two epochs, a drop of 70\% in count rate is expected due to
the decline in the detector efficiency at low energies. We attribute the difference
between the observed and expected decline to the remnant development over the 
intervening 13.8 years, indicating an X-ray flux drop of 38$\pm$5\%. No large
differences in median photon energy or hardness ratio were found between 2000 and
2013 (see also Table~\ref{tb:param} for details).

\begin{figure}[tb]
 \epsscale{1.00}
 \plotone{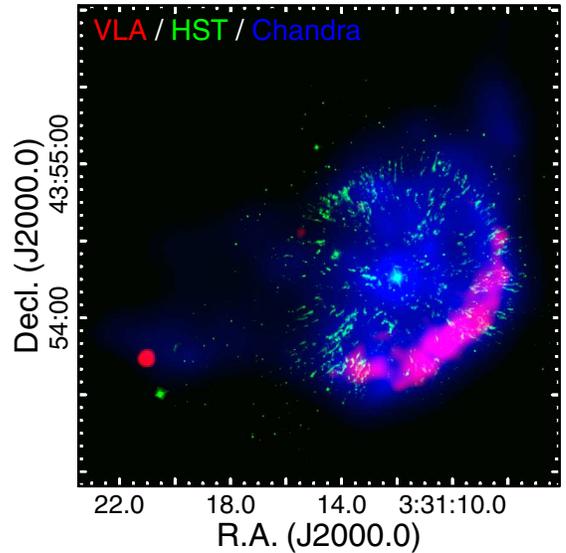}
 \caption{Comparison between radio, optical, and X-ray images of GK~Per rendered in
 red, green, and blue, respectively, on logarithmic scales tuned to clarify features
 of interest. The radio data was taken by the B-configuration \textit{VLA} in 1997
 using the L-band receiver at 1.45~GHz. The optical data was by the WFPC2 instrument
 of the \textit{HST} in 1995 using the H$\alpha+$[\ion{N}{2}] filter. The X-ray image
 is by \textit{Chandra} in 2000; the same data as the panel (a) in Figure~\ref{fg:im}.
 }\label{fg:fox}
\end{figure}

Figure~\ref{fg:im} shows 0.5--1.2~keV photon-flux images obtained by \textit{Chandra}
ACIS-S3 in both 2000 and 2013 epochs.
Exposure maps used for flat-field correction were computed at 0.6~keV, corresponding to
the median detected photon energy, using the \texttt{fluximage} software in CIAO. A
sliding cell search for source candidates (\texttt{celldetect}; \citealt{harnd1984s})
was performed, and astrometry was fixed by matching the observed X-ray positions of
three point sources with Two-Micron All Sky Survey (2MASS; \citealt{cutri2003a})
catalogue positions of their infrared counterparts: J03311201$+$4354154 (i.e., the
white dwarf binary of GK Per), J03312219$+$4356461, and J03310820$+$4357503.
The \texttt{celldetect} position estimates have precisions better than 0$\farcs$2,
and the resulting relative astrometric precision between the two epochs
based on the three reference sources remains at a similar level. The distribution of
diffuse sources was determined from a Voronoi Tessellation and Percolation algorithm
(\texttt{vtpdetect}; \citealt{ebeling1993}) applied to a combined exposure-corrected
image in which the two observations in 2000 and 2013 were merged.

We then compared the \textit{Chandra} X-ray data with the \textit{HST} optical image
and the \textit{Very Large Array} (\textit{VLA}) radio observations in
Figure~\ref{fg:fox}. The \textit{HST} data were obtained from the {\it Hubble} Legacy
Archive, and we retrieved the H$\alpha+$[\ion{N}{2}] image taken by the Wide Field
and Planetary Camera 2 (WFPC2) in 1995 --- see \citet{shara2012g} for further details.
The {\it VLA} observations were discussed by e.g.,
\citet{reynolds1984n,anupama2005,balman2005}, and the 1.45~GHz L-band image taken by
the B-configuration in 1997 was obtained from the National Radio Astronomy Observatory
\textit{VLA} Archive Survey (NVAS) system \citep{crossley2008}. We combined the data
from the different wavelength regimes using the SAOImage DS9 image analysis tool
\citep{joye2011s}, rendering radio, optical, and X-rays in red, green, and blue,
respectively. The conspicuous coincidence of the radio and X-ray emission, concentrated
at the bright rim to the southwest, was noted by \citet{balman2005}, who also presented
comparisons with optical image contours.

\begin{figure}[tb]
 \epsscale{1.00}
 \plotone{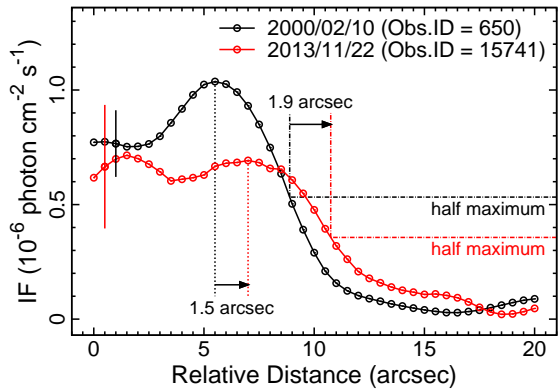}
 \caption{Radial projection profiles in the 0.5--1.2~keV band derived from the
 rectangular regions in Figure~\ref{fg:im}, smoothed by Gaussian blurs of 5~pixel radius
 at 2$\sigma$, are shown color-coded (black for 2000 and red for 2013). The vertical
 dotted lines mark the positions of the peak of the rim emission and the
 dashed-and-dotted lines are the half-maximum levels of each profile. Both indicate that
 the observed expansion of the X-ray nebula is approximately 1$\arcsec$--2$\arcsec$ in
 13.8 years. The typical 1$\sigma$ uncertainties of the data points are represented by
 the vertical bars.
 }\label{fg:proj}
\end{figure}

%%%%%%%%%%%%%%%%%%%%%%%%%%%%%%%%%%%%%%%%%%%%%%%%%%%%%%%%%%%%
\subsection{Limits on Hard X-ray Evolution}
%%%%%%%%%%%%%%%%%%%%%%%%%%%%%%%%%%%%%%%%%%%%%%%%%%%%%%%%%%%%

Unfortunately, our additional data provide no further insights into the non-thermal hard
X-ray signatures reported by \cite{balman2005} because of the background contamination
above about 1~keV. We investigated this non-thermal emission by first fitting the
extracted spectra from both epochs in the 1.2--7.0 keV range using an absorbed power-law
model and estimating the 1$\sigma$ statistical uncertainties in the power-law
normalization. Using these, we calculated count rates in the 0.5--1.2 keV range by
extrapolation. The fitted power-law slopes were approximately zero, and the resulting
upper limits to the contribution of power-law component to the count rate are 10\% and
23\% for the first and second observations, respectively. The larger limit from the
second epoch results from the larger uncertainty in the fitted slope for those data.
However, it is unlikely that the harder component has increased in intensity or
hardness while the rest of the remnant has faded significantly. To place a best
estimate on the soft X-ray power-law emission contribution, we therefore imposed
additional constraints of not allowing either the spectrum to get harder or the
normalization to increase from the first epoch to the second. We introduced a lower
limit to the power-law slope and an upper limit to the normalization corresponding to
the 1$\sigma$ limits derived from the first observation and found a 1$\sigma$ limit of
an 18\% contribution due to the harder component, although the actual contribution is
likely similar to the 10\% one of the earlier epoch. Since the power-law spectrum
would be essentially flat, it has little influence on the interpretation of the soft
X-ray data and we disregard it in the remainder of the analysis.

%%%%%%%%%%%%%%%%%%%%%%%%%%%%%%%%%%%%%%%%%%%%%%%%%%%%%%%%%%%%
\subsection{Radial Profiles and Remnant Expansion}
%%%%%%%%%%%%%%%%%%%%%%%%%%%%%%%%%%%%%%%%%%%%%%%%%%%%%%%%%%%%

In order to examine the expansion of the remnant between the two {\it Chandra}
observation epochs, the projected and summed profile of the bright southwest rim was
analyzed. A rectangular extraction region was placed on top of, and tangential to,
the rim (indicated by the red rectangles lying to the southwest in Figure~\ref{fg:im}),
and X-ray events were extracted and summed in the tangential direction such that the
resulting histogram shows the variation of integrated flux (IF) in an approximately
radial direction from the nebula center.

Figure~\ref{fg:proj} illustrates the projection profiles of these brightest regions in
the X-ray images of both epochs. Comparing the profiles at their peaks suggests an
observed expansion of 1$\farcs$5, while at half maxima linear interpolation indicates
1$\farcs$9 in 13.8 years. These two estimates are significantly larger than any
systematic error in the relative positional accuracy of the two observations. In the
following, we adopt the measurement based on the half maxima for discussing the remnant
expansion.

\begin{figure}[tb]
 \epsscale{1.00}
 \plotone{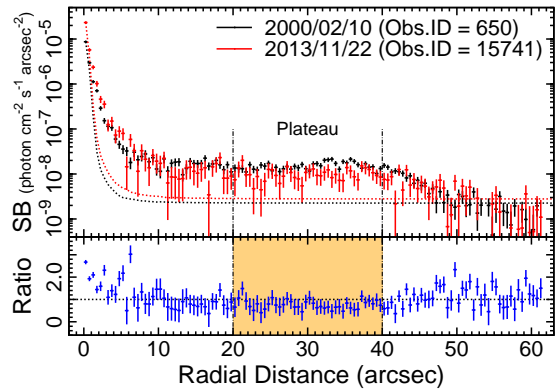}
 \caption{Comparison of background-subtracted SB radial profiles of GK~Per in 2000
 (black) and 2013 (red) in the 0.5--1.2~keV band computed from the exposure-corrected
 photon-flux images, taking into account the degradation of the quantum efficiency
 of ACIS-S3. The dashed lines show 0.6~keV PSFs that were normalized to the maxima of
 the observed peaks, and further to the minima of the background levels. The lower
 panel indicates the ratios of 2013 to 2000 data, suggesting that the SB drop was
 36$\pm$6\% in the 20$\arcsec$--40$\arcsec$ plateau area, which is indicated by the
 orange region and the dashed-and-dotted lines.
 }\label{fg:rpf}
\end{figure}

Radial surface brightness (SB) profiles derived by azimuthally summing the observed
signal as a function of radial distance from the central binary position, with a
0$\farcs$5 step size, are shown in Figure~\ref{fg:rpf}. Also shown is a 0.6~keV point
spread function (PSF) calculated using the \textit{Chandra} Ray Tracer tool
\citep{carter2003c}. Comparison of the central source counts as a function of radius
with the PSF prediction indicates the source is heavily piled up in the center. The
excess counts over the PSF model out to about 10$\arcsec$ are then expected to arise
as a result of the PSF being normalized to the peak counts, although it is possible
some of the excess could be extended emission from the remnant.

Plateaus in the SB 20$\arcsec$--40$\arcsec$ from the center indicate
regions where the nebula emission clearly dominates any signal from the central source.
Fitting a constant SB level to these segments of the radial profiles resulted in best
fits of 1.48$\pm$0.04$\times$10$^{-8}$ and
0.95$\pm$0.06$\times$10$^{-8}$~photons~cm$^{-2}$~s$^{-1}$~arcsec$^{-2}$ in 2000 and
2013, respectively, corresponding to a decay of 36$\pm$6\% in 13.8 years. The radial
profiles decay to the background levels at distances of 50$\arcsec$--60$\arcsec$.
Based on the radial profiles, the remnant signals for further spectral analysis were
extracted from annular regions between 20$\arcsec$--60$\arcsec$, avoiding potential
contamination from the PSF wings.
Background events were accumulated from the 70$\arcsec$--100$\arcsec$ annular region,
excluding the areas where diffuse emission was detected by the \texttt{vtpdetect}
algorithm (see Figure~\ref{fg:im} for details), and the background rate was computed
by normalizing to the ratio of net source and background extraction regions.

%%%%%%%%%%%%%%%%%%%%%%%%%%%%%%%%%%%%%%%%%%%%%%%%%%%%%%%%%%%%
\subsection{Azimuthal Dependence and Evolution}
%%%%%%%%%%%%%%%%%%%%%%%%%%%%%%%%%%%%%%%%%%%%%%%%%%%%%%%%%%%%

\begin{figure}[tb]
 \epsscale{1.00}
 \plotone{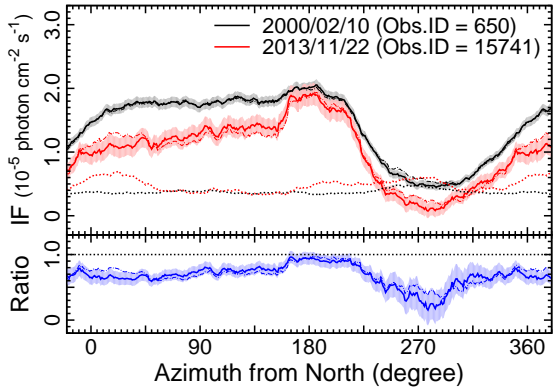}
 \caption{Comparison of 0.5--1.2~keV background-subtracted azimuthal dependence in
 integrated flux in 2000 (black) and 2013 (red). The solid and dash-dotted lines are
 the results of using different background estimates; the former used the one-eighth
 outward sectors at each azimuth angle, of which the background levels are shown by
 the dashed lines, while the latter is from the entire background annulus (see
 Figure~\ref{fg:im} for details). The 1$\sigma$ confidence ranges of the former
 estimates are illustrated by the color-coded bands. The lower panel indicates the
 ratios of 2013 to 2000 data.
 }\label{fg:az}
\end{figure}

The azimuthal dependence of the nebular emission was investigated by assessing the
source and background counts in 45$\degr$ segments of their respective annular
extraction regions and rotating these segments in the clockwise direction with a
1$\degr$ step size. The resulting IFs as a function of azimuth computed using both the
rotating background sector and just the average background in the annulus (again with
diffuse emission excluded) are illustrated in Figure~\ref{fg:az}. The results reflect
the ``by-eye'' assessment that the brightest part of the remnant is in the south and the
faintest in the east. The ratios between 2000 and 2013 curves indicate that the flux
decline is inversely related to X-ray brightness --- fainter regions are getting more
faint at a faster rate than brighter regions.  A simple average of the IF ratios is
about 70\%, with a maximum ratio of 90\% in the bright area to the south, and down to
20\% to the northwest.

%%%%%%%%%%%%%%%%%%%%%%%%%%%%%%%%%%%%%%%%%%%%%%%%%%%%%%%%%%%%
\subsection{Spectral Analysis}
%%%%%%%%%%%%%%%%%%%%%%%%%%%%%%%%%%%%%%%%%%%%%%%%%%%%%%%%%%%%

\begin{figure}[tb]
 \epsscale{1.00}
 \plotone{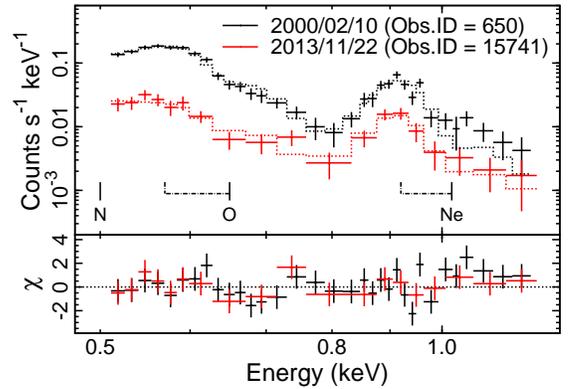}
 \caption{Background-subtracted spectra of the GK Per remnant in 2000 (black) and 2013
 (red). The best-fit models of each spectrum are shown by the dashed lines. The lower
 panel shows the residuals from the best-fit models. The energies of plausible K$\alpha$
 emission lines are labeled with solid and solid-and-dashed lines for H- and He-like
 ions, respectively.
 }\label{fg:spec}
\end{figure}

As no significant secular trends, except for background variations, were found in
the light curves constructed from the background-subtracted remnant events, spectral
analysis proceeded based on the whole of each data set. Background-subtracted X-ray
spectra from both epochs are illustrated in Figure~\ref{fg:spec}. Emission from
\ion{Ne}{9} K$\alpha$ was clearly present in both epochs, while \ion{O}{7} emission
is also possibly discernible. We applied a fitting model comprising emission from a
non-equilibrium ionization collision-dominated plasma plus photoelectric interstellar
absorption (T\"{u}ebingen-Boulder absorption; \citealt{wilms2000o}) with a
hydrogen-equivalent column density ($\nh$) fixed at 2.1$\times$10$^{21}$~cm$^{-2}$.
The absorption column was not well-constrained by the {\it Chandra} data and the
latter was estimated from the \ion{H}{1} maps of \cite{kalberla2005t} for a 1$\degr$
cone radius around the GK~Per position.
Here we experimented first with allowing the plasma abundances of C, N, O, and Ne to
vary from their solar values, as might be expected in the shell of a planetary nebula
and also in the ejecta of a classical nova. As a result no significant constraint
on the C abundance has been given from the spectral fitting. Regarding the
N emission, the best-fits and 1$\sigma$ statistical errors suggest an abundance
of 0.3--2.7 times the solar value. The reason for such poor constraints is that
the energies of the K$\alpha$ emission lines from those metals are slightly below
the useful low energy cut-off of the X-ray spectra. We confirmed that varying the
C and N abundances provides no significant difference on other parameters, and thus
the elemental abundances of the emission component were fixed to be solar, except
for O and Ne from which the energies of the K$\alpha$ lines are in the range of the
extracted spectra. For the final spectral fitting, the free parameters were then:
(1) plasma temperature ($\kt$), (2) elemental abundances of O ($\zoxy$) and Ne
($\zneo$), (3) the non-equilibrium plasma ionization timescale ($\tau$),
and (4) a volume emission measure ($\ve$). The observed fluxes ($\fx$) and
absorption-corrected luminosities ($\lx$) in the 0.5--1.2~keV energy band were
also determined. The flux decay between the 2000 and 2013 epochs determined from
these spectra amounts to 40\%.
The spectral properties and the best-fit parameters producing minima in the $\chi^{2}$
test statistic defined by \citet{gehrels1986c} are summarized in Table~\ref{tb:param}.
The results for the 2000 epoch are broadly consistent with the results of
\citet{balman2005}, and the fitted temperature for both epochs of 0.2~keV is similar to
the 2$\times$10$^{6}$~K temperature derived from the 1992 \textit{ROSAT} observation by
\citet{balman1999}. Independent spectral analysis of different regions of the nebular
was investigated but not pursued due to insufficient signal-to-noise ratio.

%%%%%%%%%%%%%%%%%%%%%%%%%%%%%%%%%%%%%%%%%%%%%%%%%%%%%%%%%%%%
\section{Discussion}\label{sc:dis}
%%%%%%%%%%%%%%%%%%%%%%%%%%%%%%%%%%%%%%%%%%%%%%%%%%%%%%%%%%%%
\subsection{Energetics}\label{ss:en}
%%%%%%%%%%%%%%%%%%%%%%%%%%%%%%%%%%%%%%%%%%%%%%%%%%%%%%%%%%%%
We start with an overview of the energetics of the explosion that lead to the GK~Per
remnant. First, assuming the angular radius $R$ $=$ 1$\arcmin$ corresponds to 0.14~pc at
a distance of 470~pc, the upper limit to the remnant volume assuming a spherical shape
is $V$ $=$ 3.4$\times$10$^{53}$~cm$^{3}$. The geometry of the emission is instead
presumably a thin shell with thickness $R/12$, as expected for a compression ratio of
4 for a strong shock, which suggests a plasma emitting volume of $V$ $=$
7.8$\times$10$^{52}$~cm$^{3}$. Here, the observed volume emission measures, $\ve$ $=$
$n_{\rm{e}}n_{\rm{i}}V$, where $n_{\rm{e}}$ and $n_{\rm{i}}$ are electron and ion number
densities, respectively, are 1--2$\times$10$^{54}$~cm$^{-3}$ at the distance 470~pc. The
larger uncertainties on the fitted $\ve$ values result from degeneracy with the
weakly-constrained Ne and O abundances. Further assuming solar abundances, the electron
to ion ratio is $n_{\rm{e}}$ $=$ 1.18$n_{\rm{i}}$, and so $n_{i}$ $\sim$ 4~cm$^{-3}$,
which corresponds to the pre-shock total number density of $n_{0}$ $\sim$ $n_{\rm{i}}/4$
$\sim$ 1~cm$^{-3}$. The swept-up mass is $M$ $\sim$ $m_{\rm{H}}n_{\rm{i}}V$ $\sim$
2$\times$10$^{-4}$~$M_{\odot}$, where $m_{\rm{H}}$ is a hydrogen mass.

If the remnant is in an adiabatic stage \citep{seaquist1989,balman2005,anupama2005}, in
which radiating energy loss is still negligible compared with the initial explosion, the
plasma state can be compared with the Sedov-Taylor solution \citep{sedov1959s}
represented by the following three equations,
\begin{small}
 \begin{eqnarray}
  R &=& 4\times10^{19} (\frac{t}{10^{4}{\rm{yr}}})^{\frac{2}{5}} (\frac{E}{10^{51}{\rm{ergs}}})^{\frac{1}{5}} (\frac{n_{0}}{1{\rm{cm}^{-3}}})^{-\frac{1}{5}} [{\rm{cm}}], \\
  v &=& 5\times10^{7} (\frac{t}{10^{4}{\rm{yr}}})^{-\frac{3}{5}} (\frac{E}{10^{51}{\rm{ergs}}})^{\frac{1}{5}} (\frac{n_{0}}{1{\rm{cm}^{-3}}})^{-\frac{1}{5}} [{\rm{\frac{cm}{s}}}], \\
  T &=& 3\times10^{6} (\frac{t}{10^{4}{\rm{yr}}})^{-\frac{6}{5}} (\frac{E}{10^{51}{\rm{ergs}}})^{\frac{2}{5}} (\frac{n_{0}}{1{\rm{cm}^{-3}}})^{-\frac{2}{5}} [{\rm{K}}].
 \end{eqnarray}
\end{small}
Here, the simultaneous equations consist of the six parameters: the radial distance to
the outer shock $R$, the blast-wave velocity $v$, the shock temperature $T$, the elapsed
time $t$, the total explosion energy $E$, and the total number density $n_{0}$. The
velocity is $v$ $=$ $(2/5)(R/t)$ $\sim$ 500~km~s$^{-1}$ with $R$ $=$ 1$\arcmin$ and $t$
$=$ 112.8 years, and similarly the temperature is $\kt$ $=$
1.8$\times$10$^{5}(R\rm{[pc]}/t\rm{[yr]})^{2}$ $\sim$ 0.28~keV. The radial distance is
then $R$ $=$ 0.31$(E\rm{[10^{51}~ergs]}/n_{0})^{1/5}$ $t\rm{[yr]}^{2/5}$~pc, which can
be converted to the explosion energy $E$ $\sim$ 2$\times$10$^{45}$~ergs using $n_{0}$
$=$ 1~cm$^{-3}$. This distance is fairly insensitive to the exact particle density
assumed, The observed electron temperature of $\kt$ $\sim$ 0.2--0.3~keV is almost
consistent with the Sedov-Taylor estimates, suggesting that the electrons and ions are
near equilibration.
All these estimates lie within typical ranges for nova explosions obtained from
both observations and theory \citep[e.g.,][]{prialnik1995e,starrfield2009e,metzger2014}.

%%%%%%%%%%%%%%%%%%%%%%%%%%%%%%%%%%%%%%%%%%%%%%%%%%%%%%%%%%%%
\subsection{Proper Motion}\label{ss:pr}
%%%%%%%%%%%%%%%%%%%%%%%%%%%%%%%%%%%%%%%%%%%%%%%%%%%%%%%%%%%%
The observed expansion of the X-ray brightest part of the remnant in the 13.8~years
separating the {\it Chandra} observations derived from the
radial-profile comparisons is 1$\farcs$9, corresponding to an expansion rate
of 0$\farcs$14~yr$^{-1}$. Assuming a distance of 470~pc, the velocity of the
outermost X-ray emitting region is 300~km~s$^{-1}$. This is slightly lower than
that the 600--1000~km~s$^{-1}$ of optical knots moving in a similar direction
\citep[][see also \citealt{Duerbeck:87}]{shara2012g,liimets2012t}, as well as the
Sedov-Taylor estimate of 500~km~s$^{-1}$. The velocity inconsistencies might imply
that the shock speed was faster than the estimate from the apparent X-ray advance and
that this measurement is not representative of the advance of the shock front itself.
Further, as we discuss in Sect.~\ref{ss:cl}, the expansion of the blast wave must have
been non-uniform.

%%%%%%%%%%%%%%%%%%%%%%%%%%%%%%%%%%%%%%%%%%%%%%%%%%%%%%%%%%%%
\subsection{Cooling or Thinning?}\label{ss:cl}
%%%%%%%%%%%%%%%%%%%%%%%%%%%%%%%%%%%%%%%%%%%%%%%%%%%%%%%%%%%%
The flux estimates based on radial profiles, the IF azimuthal dependence, and the
spectral fitting all indicate that the 0.5--1.2~keV brightness has been reduced on
average to 60--70\% of its value in 13.8 years, corresponding to a decline rate of
about 3\%~yr$^{-1}$. This is very similar to the flux decay rate of the optical knots in
the remnant of 2.6\%~yr$^{-1}$ \citep{liimets2012t}, and to the annual secular decrease
of 2.1\% in the flux density at 1.4 and 4.9 GHz between 1984 and 1997 found by
\citet{anupama2005}. There is no obvious significant X-ray spectral evolution
accompanying this decay, and all the fitted spectral model parameters for the two epochs
were statistically consistent (see Table~\ref{tb:param}).

For an emitting plasma with cosmic abundances, the observed temperature of 0.2--0.3~keV
implies a total cooling coefficient of order 10$^{-22}$--10$^{-23}$~erg~cm$^{3}$~s$^{-1}$
(e.g., \citealt{gehrels1993}). Here, the dominant cooling occurs from the line emission
of Fe \citep{boehringer1989m}. Since the shocked circumstellar environment is likely to
have a chemical composition more like that of typical planetary nebulae, the actual
cooling coefficient could differ from this value --- the tendency of the Fe depletion in
galactic planetary nebulae \citep{delgado2009i,delgado2014c} suggests that the plasma
cooling becomes less effective also in the GK~Per remnant. The cooling timescale
is therefore orders of magnitude longer than the present lifetime of the remnant. As the
temperature was almost constant during 2000--2013 as expected, the radiated energy
integrated over 13.8~years is of order 10$^{39}$--10$^{41}$~ergs or lower. This strongly
suggests that radiative cooling is negligible in comparison with the explosion energy of
order 10$^{45}$~ergs, and supports earlier conclusions that the remnant is in the
adiabatic phase \citep{seaquist1989,anupama2005,balman2005}.

We thus conclude that the most plausible cause of the 30--40\% decline in emission
between 2000--2013 is not the plasma cooling but just the thinning and expansion of the
emitting nebula. Assuming the blast velocity of 500~km~s$^{-1}$, the emitting volume
increased 20\% in 13.8~years. Were the blast wave expanding into a medium of uniform
density, the increase in swept-up mass combined with an increase in the shocked gas 
volume in the expression for the emission measure, means the X-ray emission should
increase with the cube of the remnant radius. In the Sedov phase, since $R$ $\propto$
$t^{2/5}$, the emission measure should increase with time as $\ve$ $\propto$ $t^{6/5}$.
That this is definitely not the case indicates that the medium into which the GK~Per
remnant has been recently expanding must be non-uniform.  Indeed, as discussed in
Sect.~\ref{sc:target}, the complex non-uniform nature of the circumbinary medium of
GK~Per has been evident for more than a century.  The observed decline in X-ray
emission corresponds best to the case of negligible swept-up mass during 2000--2013
and therefore a number density at the shock front in the later epoch much lower than
at earlier times. The later volume emission measure $\ve$ $=$ $n_{\rm{e}}n_{\rm{i}}V$
would then be about 80\% of the 2000 value.

The inverse relation between the X-ray brightness and the decline rate found in the
azimuthally-dependent data indicates that the currently fainter regions are fainter
because they have dimmed more rapidly, and continue to do so.  This is especially
prominent to the east, at an azimuthal angle of around 270$\degr$. These regions might
be fading faster because they are expanding more rapidly, although the X-ray data are
insufficient to test this.

Asymmetric expansion is {\em not} see in the optical knots \citep{liimets2012t} which
actually experienced a recent brightening in the east. Such divergent behaviors are
not unrealistic since optical knots are probably dominated by shocked
ejecta and we expect the X-ray emission to instead originate largely from the shocked
medium.  Comparison of the X-ray and optical emission in the multi-wavelength image
in Figure~\ref{fg:fox} reveals that some optical knots precede the X-ray emitting
shock front. Similar behavior has been seen in optical knots of the Cas~A supernova
remnant \citep{hammell2008c}. The main shell of the brightest knots is quasi-spherical,
and close inspection of the X-ray emission indicates that it is co-spatial with knots.
This indicates that the knots are likely not the same material as the X-ray emitting
plasma that has condensed and cooled, and is consistent with the knots being shocked
ejecta \citep[e.g.,][]{Bode.etal:04,liimets2012t,shara2012g}.

The most obvious departure of optical and X-ray morphology is in the X-ray ``wings''
pointing to the north and east that extend well beyond the optical knots.
\citet{balman2005} noted that the expansion rate for these wing regions must be about
2600--2800~km~s$^{-1}$ --- more than twice the rate of the optical knots and the bright
emission in the southwest --- and drew comparison to the ``cone'' of earlier ejected
material proposed in the planetary nebula scenario by \citet{seaquist1989}.  The
implication would be that we are still seeing largely the shocked medium, but that the
density experienced by the blast in these wing directions up until encountering the
denser material must have been much lower than in the southwest direction to permit the
more rapid expansion.

%%%%%%%%%%%%%%%%%%%%%%%%%%%%%%%%%%%%%%%%%%%%%%%%%%%%%%%%%%%%
\section{Conclusions}
%%%%%%%%%%%%%%%%%%%%%%%%%%%%%%%%%%%%%%%%%%%%%%%%%%%%%%%%%%%%
Comparison of {\it Chandra} X-ray imaging spectroscopy observations of the remnant of
the GK~Per classical nova event of 1901 obtained in 2000 and 2013 has revealed the
following results.
\begin{enumerate}
 \item The remnant faded significantly in the interval between the two observations,
       with a mean flux decline of 30--40\%. The rate of fading is spatially dependent,
       with more rapid fading generally coinciding with regions of lower surface
       brightness. 
 \item Comparison of X-ray and optical data confirms qualitatively the very different
       morphology of cool knots and hot gas.  The highly asymmetric X-ray remnant
       contrasts with a quasi-spherical shell of complex knots. Some optical knots
       appear to be slightly ahead of the X-ray emitting shock front, and also appear
       X-ray dark.
 \item The expansion of the remnant deduced from the difference in the bright rim of
       X-ray mission to the southwest in the two different epochs is 1$\farcs$9, or
       an expansion rate of 0$\farcs$14~yr$^{-1}$.  This corresponds to velocity of
       approximately 300~km~s$^{-1}$ for a distance to GK~Per of 470~pc.
 \item Assuming the X-ray emission arises from a relatively thin shell of shocked
       plasma, the swept-up mass is approximately 2$\times$10$^{-4}$~$M_{\odot}$ and
       the ion density is 4~cm$^{-3}$. A Sedov-Taylor solution yields an explosion
       energy of 2$\times$10$^{45}$~ergs, and electrons and ions are close to
       temperature equilibration.
 \item Since the radiative cooling time of the shock-heated gas is much longer than
       the age of the remnant, we conclude that the X-ray fading is a result of
       adiabatic expansion at a rate of about 3\% per year in the southwest rim. The
       more drastic fading in regions of low surface brightness, such as in the east
       of the remnant, is then expected to be caused by more rapid expansion into a
       lower-density medium.
 \item The general evolution of the remnant is qualitatively consistent with
       expectations based on early post-explosion photographs of pre-existing
       nebulosity, with a highly-structured and inhomogeneous shell-like morphology.
       The fading observed between 2000 and 2013 is then caused by the recent
       expansion of the remnant into regions of much lower density.  This is
       consistent with a planetary nebula origin for the pre-blast circumstellar
       material illuminated by light echoes from the explosion itself.
\end{enumerate}

%%%%%%%%%%%%%%%%%%%%%%%%%%%%%%%%%%%%%%%%%%%%%%%%%%%%%%%%%%%%
%% Acknowledgments
%%%%%%%%%%%%%%%%%%%%%%%%%%%%%%%%%%%%%%%%%%%%%%%%%%%%%%%%%%%%
\acknowledgments

We thank \textit{Chandra} for allocating a part of the telescope time in Cycle-15,
and acknowledge \textit{HST} and \textit{VLA} for their archival images. D.\,T.
thanks T.\,Yuasa, K.\,Morihana, and J.\,Ueda for valuable discussions, and
acknowledges financial support from the RIKEN/SPDR program, Grant-in-Aid for JSPS
Fellows for Research Abroad, and \textit{Chandra} grant GO4-15025X. J.\,J.\,D.
and P.\,S. were supported by NASA contract NAS8-03060 to the \textit{Chandra} X-ray
Center, and thank the Director, B.~Wilkes, for continuing advice and support.

\clearpage
%%%%%%%%%%%%%%%%%%%%%%%%%%%%%%%%%%%%%%%%%%%%%%%%%%%%%%%%%%%%
%% Bibliography
%%%%%%%%%%%%%%%%%%%%%%%%%%%%%%%%%%%%%%%%%%%%%%%%%%%%%%%%%%%%
%%%\bibliographystyle{ms}
%%%\bibliography{ms}

%%%%%%%%%%%%%%%%%%%%%%%%%%%%%%%%%%%%%%%%%%%%%%%%%%%%%%%%%%%%
\end{document}